\documentclass{emulateapj}   

\usepackage{wrapfig}
\usepackage{graphicx}
\usepackage{epsfig}
\usepackage{epstopdf}

\usepackage{natbib}
\newcommand{\Mach}{\mathcal{M}}

\newcommand{\etal}{et~al.~}

\begin{document}

\title {Cosmological Shocks in Adaptive Mesh Refinement Simulations and the Acceleration of Cosmic Rays}
\author{Samuel~W.~Skillman\altaffilmark{1,2}, Brian~W.~O'Shea\altaffilmark{2,3}, Eric~J.~Hallman\altaffilmark{1,4}, Jack~O.~Burns\altaffilmark{1}, Michael~L.~Norman\altaffilmark{5}}

\altaffiltext{1}{Center for Astrophysics and Space Astronomy, Department of Astrophysical \& Planetary Science, University of Colorado, Boulder, CO 80309}

\altaffiltext{2}{Theoretical Astrophysics (T-6), Los Alamos National Laboratory, Los Alamos, NM 87545}

\altaffiltext{3}{Department of Physics \& Astronomy, Michigan State University, East Lansing, MI, 48824}

\altaffiltext{4}{National Science Foundation Astronomy and Astrophysics Postdoctoral Fellow}

\altaffiltext{5}{Center for Astrophysics and Space Sciences, University of California-San Diego, 9500 Gilman Drive, La Jolla, CA 92093}

\email{samuel.skillman@colorado.edu}

\begin{abstract}
We present new results characterizing cosmological shocks within adaptive mesh refinement N-Body/hydrodynamic simulations that are used to predict non-thermal components of large-scale structure.  This represents the first study of shocks using adaptive mesh refinement.  We propose a modified algorithm for finding shocks from those used on unigrid simulations that reduces the shock frequency of low Mach number shocks by a factor of $\sim3$.  We then apply our new technique to a large, $(512~Mpc/h)^3$, cosmological volume and study the shock Mach number~($\Mach$) distribution as a function of pre-shock temperature, density, and redshift.  Because of the large volume of the simulation, we have superb statistics that results from having thousands of galaxy clusters.  We find that the Mach number evolution can be interpreted as a method to visualize large-scale structure formation. Shocks with $\Mach<5$ typically trace mergers and complex flows, while $5<\Mach<20$ and $\Mach>20$ generally follow accretion onto filaments and galaxy clusters, respectively.  By applying results from nonlinear diffusive shock acceleration models using the first-order Fermi process, we calculate the amount of kinetic energy that is converted into cosmic ray protons.  The acceleration of cosmic ray protons is large enough that in order to use galaxy clusters as cosmological probes, the dynamic response of the gas to the cosmic rays must be included in future numerical simulations.        
\end{abstract}

\section{Introduction}\label{Intro}

What determines the thermal history of galaxy clusters?  On large scales, it is governed by the in-fall of material onto dark matter halos and the conversion of gravitational potential energy into thermal energy.  This process occurs through the heating of the gas via strong accretion shocks surrounding galaxy clusters and filaments \citep{Ryu:2003aa, Miniati:2001aa, Pfrommer:2006aa, Pavlidou:2006aa}.  Once inside collapsed structures, complex flows associated with the merging of subhalos continue to create moderate-strength shocks that allow the halos to virialize.  Because of this, shocks encode information about structure formation and its thermal effects on the gas.

Cosmological shocks affect three important realms of structure formation and leave feedback on the surrounding medium.  First, shocks thermalize the incoming gas, providing much of the pressure support in baryons.  This process is the basis upon which clusters are able to virialize.  Additionally, the thermalization of gas at the standing accretion shocks surrounding large-scale filaments produces the warm-hot intercluster medium (WHIM) at temperatures of $10^5 K - 10^7 K$ \citep{Dave:1999aa,Cen:1999aa}.  The history of the mass flux through these shocks describes the evolution of gas in the WHIM phase \citep{Pfrommer:2008aa}.  

Second, the strength of the outer accretion shocks onto halos, characterized by the Mach number, is dependent upon the mass of the gravitating object.  This is because the higher mass generates larger acceleration and velocity in the diffuse gas while the sound speed of the upstream gas is uniform for previously unshocked gas, corresponding to a temperature of $\approx10^4K$.  This temperature floor is created by the reionization from stars.  Thus, the Mach number of accretion shocks can be used as an independent measure of cluster mass.  This could conceivably be a powerful new tool for cluster mass estimation if we are able to observe the accretion shock with radio observations \citep[e.g.][]{Giacintucci:2008aa}.

Finally, because these shocks are collisionless features whose interactions in the hot plasma are mediated by electromagnetic fields, it is possible for a portion of the thermal distribution of particles to be accelerated and transformed into non-thermal populations of cosmic rays (CRs) through the process of diffusive shock acceleration \citep[DSA; e.g.][]{Drury:1986aa,Blandford:1987aa}.  This process results in a fraction of the kinetic energy of shocking gas being converted into both thermal and non-thermal components \citep{Kang:2002aa,Kang:2005aa,Kang:2007aa}.  The cosmic ray electron populations are likely sources of radio halos and radio relics in galaxy clusters \citep{Pfrommer:2008aa,Kim:1989aa, Giovannini:1993aa}, while the cosmic ray protons may be sources of $\gamma$-ray emission through their interactions with gas protons.  If a significant portion of the gas pressure resides in cosmic rays, then it will likely affect gas mass fraction estimates as well as the assumption of hydrostatic equilibrium.  Because of the importance of these mass estimates in measuring dark energy, we must include the underlying physics in order to perform precision cosmology \citep{Allen:2008aa}.

To date, studies of cosmological shocks have included observational, theoretical, and numerical techniques.  Observationally, the majority of the work surrounding cosmic shocks are related to radio relics, of which only a few have been studied in depth \citep[e.g.][]{Rottgering:1997aa, Clarke:2006aa, Orru:2007aa}.  Using the spectral index of the non-thermal particle distribution, we can infer a Mach number if the acceleration is due to first order Fermi acceleration \citep{Giacintucci:2008aa}.  Additionally, GLAST will begin observing $\gamma$-rays and will likely see signatures from galaxy clusters due to hadronic cosmic ray interactions with pions \citep{Pfrommer:2008ab}.   

On the theoretical side, the majority of analyses are based upon manipulating the Press-Schechter formalism \citep{Press:1974aa,Sheth:1999aa} to deduce first the mass function of accreting objects and then their interactions with infalling material.  \citet{Pavlidou:2006aa} extended these analyses to calculate the energy and mass flux through accretion shocks.  Furthermore, several analytical attempts have been made to describe merger shocks, including those by \citet{Fujita:2001aa} and \citet{Gabici:2003aa,Gabici:2003ab}.  However, it is quite difficult to account for the complex morphologies that arise during structure formation using purely analytical frameworks.  For this reason, multiple numerical techniques have been developed using hydrodynamical simulations.  

There have been numerical studies of shocks using both Eulerian ``single-grid" codes \citep[e.g.][]{Ryu:2003aa, Kang:2007aa} as well as smoothed particle hydrodynamics (SPH) codes \citep{Pfrommer:2006aa,Pfrommer:2007aa,Pfrommer:2008aa,Pfrommer:2008ab}.  There are advantages and disadvantages of both methods.  In previous work using grid-based codes, shocks were analyzed during post-processing by examining temperature jumps for a given point in time \citep{Ryu:2003aa}.  However, it was impossible to cover the spatial dynamic range needed to describe both the complex flow within halos and their coupling to large scale structures because of the use of a uniform grid.  Therefore, even the largest simulations, with $1024^3$ cells in a $100 h^{-1} Mpc$ volume, has a comoving spatial resolution of only $97.7 h^{-1} kpc$ \citep{Ryu:2003aa}.  There have been recent attempts at prescribing hybrid models to study turbulent generation, but full resolution convergence of the results have still not yet been achieved \citep{Ryu:2008aa}.  The advantage of a grid code is its superb shock-capturing algorithms that do not rely on the use of artificial viscosity when using higher-order methods \citep{OShea:2005ab}.  

In contrast, SPH codes are implicitly adaptive in space due to the Lagrangian nature of the method, e.g. high density regions are resolved by a larger number of particles than low-density regions. This approach conserves hydrodynamic quantities exactly when they are advected with the flow. However, because the properties of the gas are determined by a weighted average over neighboring particles, formally discontinuous shocks are spread over a length determined by the smoothing length. Additionally, SPH relies on artificial viscosity to dissipate flows and produce the correct amount of entropy. Because of these restrictions, \citet{Pfrommer:2006aa} developed a method that is able to identify shocks by examining the time-evolution of the entropy of individual SPH particles. Comparing the instantaneous entropy injection rate to the characteristic time it takes a particle to cross the broadened shock surface, they are able to identify and calculate the instantaneous Mach number of shocks while remembering the pre-shock conditions. Therefore, the analysis can be performed on-the-fly and shock quantities were traced along with the usual hydrodynamic properties.  These calculations use calibrations against ``shock tube" simulations to derive the correct relationship between entropy injection rate and Mach number, which may vary with respect to different artificial viscosity implementations.  

To address all of the problems listed above, we have developed a novel numerical algorithm capable of detecting and identifying shocks in the 3-D adaptive mesh refinement (AMR) grid-based code, $Enzo$.  The use of AMR allows us to analyze unprecedented dynamic ranges with an advanced hydrodynamic code that is able to capture shocks exceedingly well.  In this work we explore simulations with dynamic range of up to $2^{16}=65,536$, but we are not limited from going further in future work.  Because of the complexity of the structure of AMR simulations, it was necessary to develop several new numerical algorithms to identify the shocks.  This shock-finding analysis algorithm will be presented and compared to previous methods \citep[e.g.][]{Ryu:2003aa, Pfrommer:2006aa}.  In order to validate and quantify the robustness of our method, we carry out a resolution study that includes both mass and spatial resolutions that vary by factors of $16$ and $64$, respectively.

We also propose a new method of characterizing shocks by their pre-shock overdensity and temperature.  This then allows analysis that goes beyond the traditional internal vs. external (of filaments/clusters) shocks classification suggested by \citet{Ryu:2003aa}.  By refining the temperature and density ranges examined, we are able to identify shocks that reside in voids, filaments, and halos.  Additionally, by using temperature cuts, we can identify the population of gas that is being shocked into the warm-hot intercluster medium (WHIM).  

After calculating the shock structure in a given simulation, we are able to compute the amount of shock kinetic energy that is transferred to high-energy cosmic ray protons through diffusive shock acceleration.  While the surface area of the large scale structure shocks are dominated by low pre-shock temperature and density, the bulk of the cosmic ray energy generation occurs in the centers of collapsed structures.  Since stronger shocks will produce harder spectra \citep{Blandford:1987aa}, we expect that the strong accretion shocks could be the source of high energy cosmic rays.  

In Section 2 we describe the numerical methods used for both the cosmological simulations as well as the analysis of the shock-finding process.  In Section 3, we compare our algorithm to that of \citet{Ryu:2003aa} and test it using 3-D ``shock tube" tests.  Section 4 contains the main results of analyzing a large, $(512~Mpc/h)^3$, cosmological simulation with a peak spatial resolution of $7.8\ kpc/h$. Section 5 describes the effects of spatial and mass resolution on the shock populations and cosmic ray acceleration.  In Section 6, we discuss the limitations of our analysis, and in Section 7 we summarize our findings and discuss potential future directions.

\section{Methodology}

\subsection{The Enzo Code}
All simulations were run using the Enzo cosmology code~\citep{Bryan:1997aa,Bryan:1997ab, Norman:1999aa, OShea:2004aa}.  While a full description can be found in the cited papers, we will review the key aspects that are of importance to this work.  

Enzo is a block-structured adaptive mesh refinement \citep[AMR;][]{Berger:1989aa} code that couples an Eulerian hydrodynamics method that follows the gas dynamics with an N-Body particle mesh (PM) solver \citep{Efstathiou:1985aa, Hockney:1988aa} to follow the dark matter component.  Enzo implements two hydrodynamic solvers.  The first is a piecewise parabolic method \citep[PPM;][]{Woodward:1984aa} with cosmological modifications by \citet{Bryan:1995aa}.  The second is the method from the ZEUS magnetohydrodynamics code \citep{Stone:1992aa,Stone:1992ab}.  In this work we restrict ourselves to the PPM method because of its superior shock-capturing ability and lack of artificial viscosity.

The AMR scheme within Enzo is handled by partitioning the simulation volume into 3D rectangular solid grids.  Each of these grids contain a number of grid cells that set the spatial scale on which the hydrodynamics is solved.  If a region of cells within a grid is determined to require higher resolution, as judged by a number of refinement criteria including gas/dark matter overdensity, minimum resolution of the Jeans length, local gradients of density, pressure, or energy, shocks, or cooling time, then a minimum enclosing volume around those cells is created at the appropriate level of refinement.  These newly created ``child grids" can then themselves recursively become ``parent grids" to yet another more highly refined region.  This recursive nature does not set any restrictions to the number of grids or level of refinement.  However, because structure formation leads to an enormous dynamic range we are limited by available computational resources, a maximum level of refinement $l_{max}$ is instituted.

In addition to being adaptive in space, Enzo implements an adaptive time stepping algorithm.  All grids on a given level are given advanced simultaneously with a maximum timestep such that the Courant condition is satisfied by all the cells on that level.  This results in a hierarchy of timesteps: a parent grid on level $l$ is advanced $\Delta t(l)$, and then its subgrid(s) on level $l+1$ are advance by one or more timesteps until they reach the same physical time as their parent grid.  At this point, flux information is exchanged from child to parent grid in order to provide a more accurate solution to the hydrodynamics on the parent grid.  This procedure is done recursively until all grids are at the same physical time as the root grid, at which point the process is repeated until the stopping point of the calculation is reached.

\subsection{Shock-Finding Algorithm}

The bulk of our analysis relies on accurately identifying and quantifying the strength of shocks.  After finding a shock, we would like to calculate its Mach number, which characterizes the strength of the shock.  There are several methods that can be used in order to calculate the Mach number, including density, temperature, velocity, or entropy jumps across the shock.  As in \citet{Ryu:2003aa}, we use the Rankine-Hugoniot temperature jump conditions to calculate the Mach number.  The temperature jump is preferable to density because it is more sensitive to Mach number, whereas the density jump quickly asymptotes for strong shocks.  The Mach number is solved for by
\begin{equation}
\frac{T_2}{T_1} = \frac{(5\Mach^2 - 1)(\Mach^2 + 3)}{16\Mach^2},
\end{equation}
where $T_2$ and $T_1$ are the post-shock~(downstream) and pre-shock~(upstream) temperatures, respectively.  $\Mach$ is the upstream Mach number. 

 A cell is determined to have a shock if it meets the following requirements: 
\begin{eqnarray}
\nabla \cdot \vec{v} < 0 \\
\nabla T \cdot \nabla S > 0\\
T_2 > T_1 \\
\rho_2 > \rho_1,
\end{eqnarray}
where $\vec{v}$ is the velocity field, $T$ is the temperature, $\rho$ is the density, and $S=T/\rho^{\gamma-1}$ is the entropy.  In our analysis, as in \citet{Ryu:2003aa}, we have set a minimum preshock temperature of $T=10^4K$ since the low-density gas in our cosmological simulations is assumed to be ionized~(a reasonable assumption at $z \lesssim 6$).  Therefore, any time the pre-shock temperature is lower than $10^4K$, the Mach number is calculated from the ratio of the post-shock temperature to $10^4K$.  This introduction of a temperature floor prevents us from drastically overestimating the accretion shock strength in adiabatic simulations.  Future work will incorporate a self-consistent UV ionizing background radiation. 

Now the task is to identify all of the shocks and their corresponding Mach numbers.  The method that has been previously used in unigrid simulations \citep[e.g.][]{Ryu:2003aa} is to loop through rows of cells along each of the coordinate axes and identify 1-D shock structures in each direction.  Contiguous cells that meet the requirements above in Eqs. 2-5 are then combined into a single shock structure with the cell of maximum convergence marked at the center.  The pre- and post-shock cells are identified as first cells outside of the shock structure.  If a center is marked in more than one direction, the maximum calculated Mach number of the three possible is taken to be the true Mach number.  Because of this, we would expect errors to arise when examining shocks whose direction of motion is not oriented along a coordinate axis.  To address this issue, we have designed an algorithm that does not suffer from this limitation. 

\begin{figure}[tp]
\centering
\includegraphics[width=0.4\textwidth]{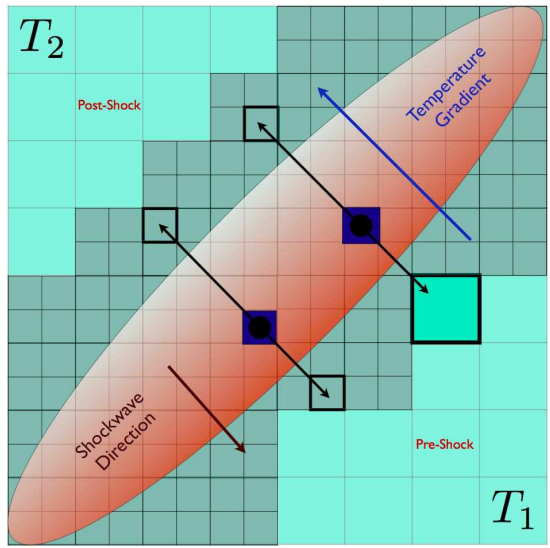}
\caption{A 2-D cartoon of our shock analysis algorithm.  The shock centers 
are shown as dark blue cells, while the pre- and post-shock cells are outlined in 
thick black.  The AMR resolution level is seen by varying grid-cell sizes.}\label{fig:2Dmethod}
\end{figure}

In our method, we first determine the direction of shock propagation from the local temperature gradients, making the assumption that the shock-induced temperature gradient overwhelms the background temperature gradient.  We then search the cells along the temperature gradient for the pre- and post-shock cells.  If we find a neighboring cell to have a more convergent flow, that cell is marked as the center and we move outwards from it.  This guarantees that the analysis is anchored to the center of the shock.  Once the furthest pre- and post-shock cells are found, the temperatures are taken and the Mach number is calculated from Equation 1.  A two-dimensional analog of this process is shown in Figure \ref{fig:2Dmethod}.  Because our algorithm is not confined to operate along coordinate axes, the calculated Mach numbers provide a more accurate description of the underlying shock properties than the \citet{Ryu:2003aa} method.  

Specifically, in situations where there are weak shocks or complex flow velocities, using the coordinate-split approach may allow for an excess of shocks since the direction of the shockwave is not taken into account.  Our method picks a single direction that a shock could be propagating, given a specific temperature gradient.  \citet{Ryu:2003aa} claim that their shocks are spread out over 2-3 cells, of which one is marked as the center.  For the other 1-2 cells, the coordinate split approach may mis-identify these other 1-2 cells as low Mach number shocks due to normal temperature gradients.   This would lead to an over-prediction of low Mach number shocks.

This process is further complicated by the use of adaptive mesh refinement (AMR), primarily because neighboring cells are not necessarily at the same level of refinement.  This occurs most often at the site of accretion shocks onto halos and filaments where the density gradient, upon which the refinement criteria are based, is largest.  Therefore, knowledge of the grid hierarchy must be used.  We incorporate this into our algorithm and allow for a neighboring cell to be any cell at the same or lower level of refinement.  We do not allow the algorithm to search for neighbors at higher levels (smaller grid cells) since one cell will have multiple neighbors.  If this situation occurs, we use the neighboring cell on the same level.  Because of this requirement, we must perform our analysis on the most highly refined grids first, and move to progressively coarser levels of resolution.


\subsection{Cosmic Ray Acceleration Models}

\begin{figure}[tp]
\centering
\includegraphics[width=0.4\textwidth]{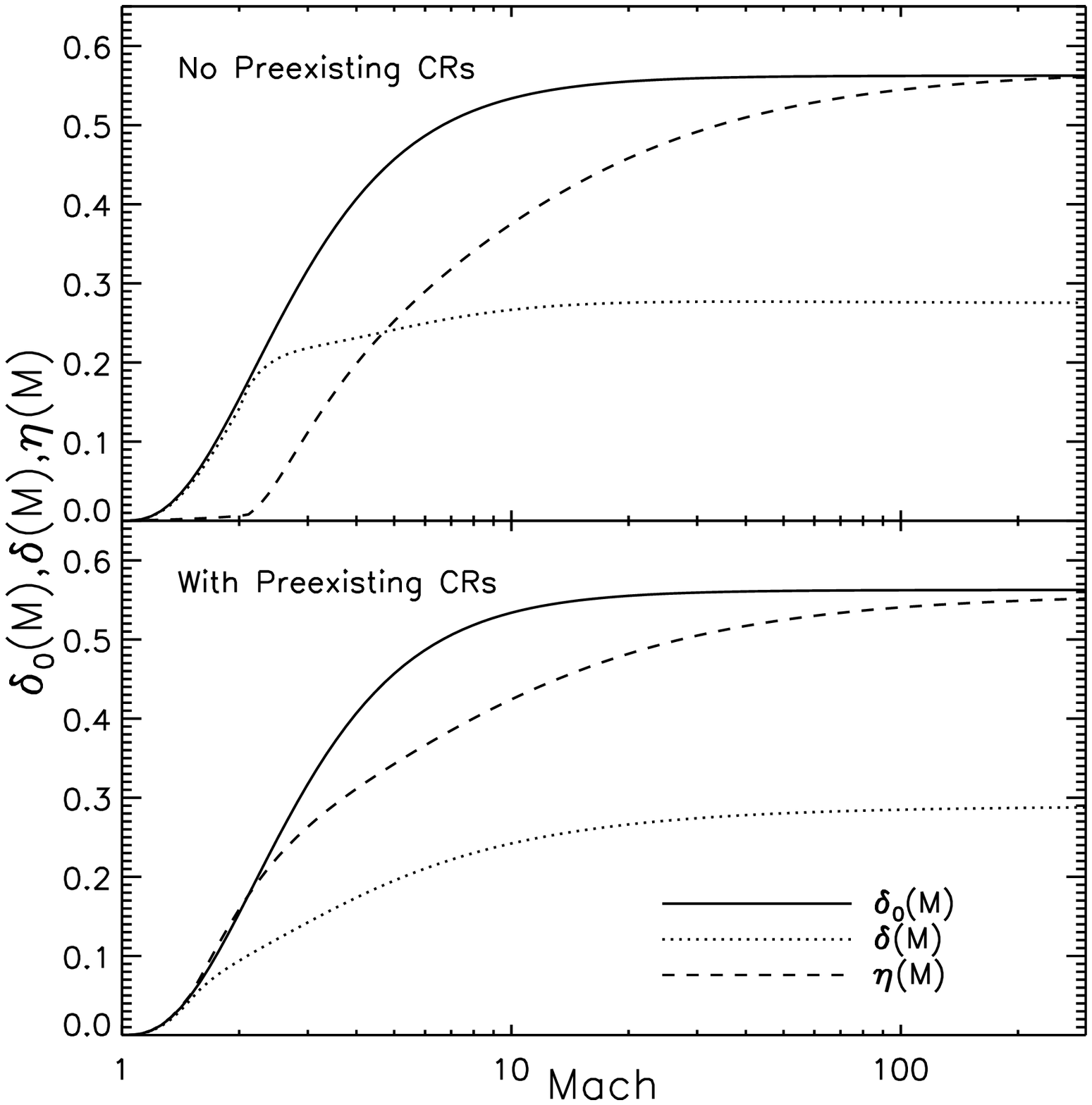}
\caption{Fractional efficiency of gas thermalization and cosmic ray acceleration, from the models by \citet{Kang:2007aa}.  $\delta_0(\Mach)$ is the gas thermalization fraction expected from the Rankine-Hugoniot jump conditions.  $\delta(\Mach)$ and $\eta(\Mach)$ are the gas thermalization fraction and cosmic ray acceleration fraction,respectively, from the non-linear calculations of \citet{Kang:2007aa}.  The two panels show results assuming different compositions of the pre-shock plasma.  Top: thermal gas with no cosmic ray population.  Bottom:  thermal gas with a pre-exisiting cosmic ray population having $P_{CR}/P_{gas} = 0.3$, where $P_{CR}$ and $P_g$ are the cosmic ray and total gas pressure, respectively.}  
\label{fig:KangJonesEff}
\end{figure}

Following the method proposed in \citet{Ryu:2003aa}, we now seek to determine the amount of kinetic energy that is converted into heating of the gas and accelerating cosmic rays.  We begin with calculating the total kinetic energy flux through a shock surface.  The kinetic energy flux associated with a mass flux of $\rho_1\Mach c_s$ is:
\begin{eqnarray}
f_{KE} = \frac{1}{2} \rho_1 \left(\Mach c_s\right)^3, 
\end{eqnarray}
where $\rho_1$ is the pre-shock density and $c_s$ is the sounds speed in the pre-shock gas.  From this total incoming kinetic energy flux, a fraction will be used in the thermalization of the gas and the acceleration of cosmic rays.  In keeping with \cite{Ryu:2003aa}, we will denote the amount of energy per unit time used to heat the gas and accelerate cosmic rays as $f_{TE}$ and $f_{CR}$, respectively.  In the case of a purely hydrodynamical shock without the inclusion of cosmic ray feedback, the fractional thermalization $\delta_0(\Mach)$ can be determined by the Rankine-Hugoniot jump conditions, 
\begin{eqnarray}
\delta_0(\Mach) = \frac{ \left[e_{TE,2} - e_{TE,1}\left(\frac{\rho_2}{\rho_1}\right)^\gamma\right] v_2 }{\frac{1}{2}\rho_1v_1^3},
\end{eqnarray}
where $e_{TE,1}$ and $e_{TE,2}$ are the thermal energy densities in the pre and post-shock regions, respectively.  

\begin{deluxetable*}{ccccccccc}[tp]
\tablecolumns{9}
\tablewidth{1.0\textwidth}
\tablecaption{Simulation Parameters}
\tablehead{
\colhead{Name} & \colhead{$L_{box}$} & \colhead{$\Delta x_{RG}$} & \colhead{$l_{max}$} & \colhead{$M_{dm}$} & \colhead{$\Delta x_{max}$} & \colhead{$\Omega_b$} & \colhead{$\Omega_m$} & \colhead{$\sigma_8$}
}
\startdata
 $ryu1024$ 	& $100$ & $97.7$ & 0 & $5.877\times 10^{7} $ & 97.7 & 0.043 & 0.27 & 0.8 \\
 $SF\ Light~Cone$ 	& $512$ & $1$ & 7 & $7.228 \times 10^{10} $ & 7.8 & 0.04 & 0.3 & 0.9  \\
 $m1\_l8$		& $256$ & $1$ & 8 & $ 6.224 \times 10^{10} $ & 3.9 & 0.0441 & 0.268  & 0.9\\
 $m1\_l6$		& $256$ & $1$ & 6 & $ 6.224 \times 10^{10} $ & 15.6 & 0.0441 & 0.268 & 0.9  \\
 $m1\_l4$		& $256$ & $1$ & 4 & $ 6.224 \times 10^{10} $ & 62.4 & 0.0441 & 0.268 & 0.9 \\
 $m4\_l8$		& $256$ & $500$ & 8 & $ 7.781\times 10^{9} $ & 3.9 & 0.0441 & 0.268 & 0.9 \\
 $m4\_l6$		& $256$ & $500$ & 6 & $ 7.781\times 10^{9} $ & 15.6 & 0.0441 & 0.268 & 0.9 \\
 $m4\_l4$		& $256$ & $500$ & 4 & $ 7.781\times 10^{9} $ & 62.4 & 0.0441 & 0.268 & 0.9 \\
 $m16\_l8$	& $256$ & $250$ & 8 & $ 9.726\times 10^{8} $ & 3.9 & 0.0441 & 0.268 & 0.9 \\
 $m16\_l6$	& $256$ & $250$ & 6 & $ 9.726\times 10^{8} $ & 15.6 & 0.0441 & 0.268 & 0.9 \\
 $m16\_l4$	& $256$ & $250$ & 4 & $ 9.726\times 10^{8} $ & 62.4 & 0.0441 & 0.268 & 0.9 \\ 
\enddata
\tablecomments{ $L_{box}$ is the simulation box size in comoving Mpc/h. $\Delta x_{RG}$ is the effective root grid resolution (the m4 and m16 series of calculations use one and two static nested grids, respectively). $l_{max}$ is the maximum level of AMR allowed in the simulation.  $M_{dm}$ is the dark matter particle mass (in the static nested grids for the m4 and m16 series of runs) in $M_\odot/h$.  $\Delta x_{max}$ is the peak spatial resolution in comoving $kpc/h$. $\Omega_b$ and $\Omega_m$ are the fractional densities of baryons and matter compared to the critical density~($\Omega_\Lambda \equiv 1-\Omega_m$ in all simulations, so $\Omega_0=1$).  $\sigma_8$ is the power spectrum normalization of the mass fluctuation in a comoving 8 Mpc sphere.}
\label{tab:simparams}
\end{deluxetable*}

With the inclusion of cosmic rays, there is no simple analytical form for the fractional thermalization of the gas, which depends on magnetic field orientation, turbulence, and the pre-shock cosmic ray population.  Instead, we adopt the results of 1D diffusive shock acceleration (DSA) simulations by \citet{Kang:2007aa}.  The time-asymptotic values of the fractional thermalization, $\delta(\Mach)=f_{TE}/f_{KE}$, and fractional CR acceleration, $\eta(\Mach)=f_{CR}/f_{KE}$, were found to be nearly self-similar for the temperatures and shock velocities considered.  These simulations also accounted for whether or not the pre-shock medium had preexisting cosmic rays.  With a preexisting cosmic ray population, the fractional energy deposited into CRs increases dramatically at low Mach numbers because it is much easier to accelerate an existing power-law distribution of particles than a thermal distribution of particles.  Shown in Figure \ref{fig:KangJonesEff} are the results of the Kang \& Jones DSA simulations for a population with no preexisting cosmic rays and one in which CRs existed initially with $P_{CR}/P_g \approx 0.3$, where $P_{CR}$ and $P_g$ are the cosmic ray and total gas pressure, respectively.  The sum of $\delta(\Mach)$, $\eta(\Mach)$, and the remaining fraction of kinetic energy in the gas is equal to one, conserving energy.

We do not track the cosmic ray population in our simulations at present, and as a result we are unable to constrain the amount of preexisting CRs in the pre-shock medium.  Therefore, we can think of our results from the two scenarios shown in Figure \ref{fig:KangJonesEff} as bracketing the likely range of efficiencies.  Additionally, these efficiency models are only valid for situations where the shock normal is parallel to the magnetic field.  Any deviation from these ideal conditions will likely reduce the efficiency of cosmic ray acceleration \citep{Kang:2007aa}, so one can view the results described later in this paper as upper limits on cosmic ray injection efficiency.


\subsection{Simulations}

We constructed three distinct sets of cosmological simulations for this project.  A summary of some of the simulation parameters is given in Table \ref{tab:simparams}.  First, we have a simulation that was devised as an analog of the unigrid numerical simulation by Ryu \etal (2003).  For this simulation, we used identical cosmological parameters to \citet{Ryu:2003aa} in order to provide a reference simulation to compare our new shock-finding method with previous work.  The cosmological parameters for this simulation, $ryu1024$, are: $\Omega_{BM} = 0.043$, $\Omega_{DM} = 0.227$, $\Omega_{\Lambda}=0.73$, $h= H_0/(100\ km\ s^{-1} Mpc^{-1})=0.7$, and $\sigma_8 = 0.8$, which are broadly consistent with WMAP Year 5 results \citep{Komatsu:2008aa}.  The comoving size of the simulation volume is $(100~Mpc/h)^3$ and is discretized into $1024^3$ cells, giving a comoving spatial resolution of $97.7 kpc/h$.  It also employs $512^3$ dark matter particles with a $1024^3$ grid. In order to reproduce the Ryu \etal results as closely possible, we did not use AMR techniques.   The simulation was initialized with an Eisenstein \& Hu (1999) power spectrum with a spectral index of $n=1.0$ at z=99 and the simulation states were output in 20 times between $z=10$ and $z=0$.  The analysis of this simulation is described in Section 3.2.

Our main results in this work focus on the analysis of a $(512~Mpc/h)^3$ volume that utilizes a $512^3$ root grid and up to 7 levels of AMR. It is referred to as the ``Santa Fe Light Cone," and has been previously described by \citet{Hallman:2007aa}.  This simulation has a peak spatial resolution of $~7.8 kpc/h$ and a dynamic range of $65,536$.  The cosmological parameters used were: $\Omega_M=0.3$, $\Omega_{BM} = 0.04$, $\Omega_{CDM} = 0.26$ , $\Omega_{\Lambda}=0.7$, $h= H_0/(100\ km\ s^{-1} Mpc^{-1})=0.7$, and $\sigma_8 = 0.9$, and employs a Eisenstein \& Hu (1999) power spectrum with a spectral index $n=1.0$.  Cells are refined whenever the baryon or dark matter density increased by a factor of 8 beyond the previous level.  Because the simulation then refines by a factor of 8 in volume, the average mass per grid cell stays roughly constant.  The simulation was initialized at $z=99$ and was run to $z=0$.  The analysis of this simulation is described in Section 4.

In order to study the effects of spatial and dark matter mass resolution, we have performed a suite of simulations that vary these factors, and illustrate their results in Section 5.  These simulations are known as ``nested grid" simulations.  An initial cosmological simulation is run at low resolution.  The most massive halo at $z=0$ is found and the simulation is re-centered at the final location.  The simulation is then re-run, while only adaptively refining a region that bounds all dark matter particles that eventually are inside the most massive halo.  Therefore, the focus of the simulation is only on the inner portion of the initial volume.  With this capability, we are able to modify the root grid and peak spatial resolution for this subvolume and study their direct effects on the evolution of a single cluster.  In our simulations, we initialize a $(256~Mpc/h)^3$ volume with $256^3$ root grid cells.  From that, we only adaptively refine in a $(32~Mpc/h)^3$ subvolume.  Within the subvolume, we add up to two static nested grids, with more highly refined dark matter particles and gas cells.  A list of all simulation parameters used is given in Table \ref{tab:simparams}.  The cosmological parameters used are: $\Omega_M = 0.268$, $\Omega_{BM} = 0.0441$, $\Omega_{CDM} = 0.2239$ , $\Omega_{\Lambda}=0.732$, $h= H_0/(100\ km\ s^{-1} Mpc^{-1})=0.704$, and $\sigma_8 = 0.9$.  These parameters are the WMAP year 3 parameters \citep{Spergel:2003aa} but with a somewhat higher $\sigma_8$.\newline

\section{Validation of Shock-Finding Method}
\subsection{Shock Tube Test}
In order to verify that our shock-finding algorithm is accurate, we have performed a suite of 3-D AMR shock tube tests.  In these tests we have varied the Mach number as well as orientation with respect to the coordinate axis.  The setup of this test problem is described in \citet{Mihalas:1984aa}.  It consists of a stationary, uniform pre-shock medium.  The shock is then introduced via boundary conditions that match the Rankine-Hugoniot Jump conditions for a given Mach number.  The volume is then allowed to adaptively refine up to 2 levels, using shocks as a criteria for refinement.   We have chosen to adaptively refine based on shock locations (i.e. strong pressure jumps) instead of density because this should introduce a complicated AMR topology in order to test the robustness of our algorithm.  This forces us to traverse different levels of refinement for pre- and post-shock quantities.

In order to change the direction of shock propagation, we change the time at which a given boundary cell changes from uniform to ``shocked."  Using this procedure, we vary the shock propagation vector over both $\theta$ and $\phi$, which are angles off of the x-z and x-y planes, respectively.  In addition to the three on-axis scenarios, we vary $\theta$ and $\phi$ over all permutations of the angles $0,\ \pi/8,\ \pi/6$, and $\pi/4$.  For each shock propagation direction, we then vary the input Mach number over $\Mach=$2, 5, 30, and 100.  

\begin{figure}[tp]
\centering
\includegraphics[width=0.4\textwidth]{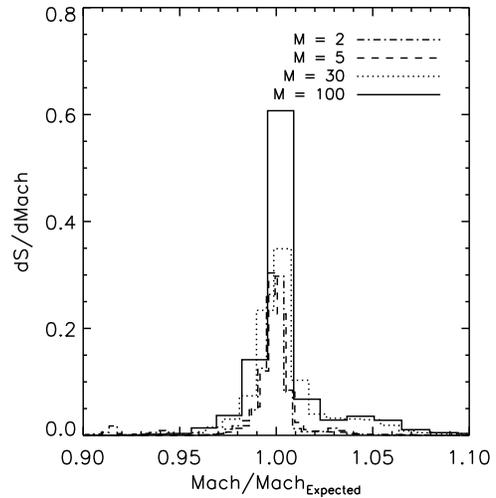}
\caption{Distributions of the ratio of calculated Mach number to expected mach number from off-axis 3D AMR shock tube test problems.  Shock surface area, S, distributions were averaged over all orientations for the each individual Mach number, and normalized so that the area under the curve is 1.  Varying lines correspond to Mach numbers of 2~(dash-dotted), 5~(dashed), 30~(dotted), and 100~(solid).  Sample standard deviations from 1.0 are all less than 0.06}\label{fig:SA_mach}
\end{figure}

The general result from this study is that our shock-finding algorithm is very accurate.  As shown in Figure \ref{fig:SA_mach}, if we make a histogram of the ratio of calculated Mach number to expected Mach number, and normalize it so that the area under the curve is equal to 1, the result is both accurate and precise.  In Figure \ref{fig:SA_mach}, we created the histogram by summing over all orientations of the shock of a given Mach number.  As one can see, the peak is centered around 1.0, with an average sample standard deviation of less than 0.06.  We have examined the average Mach number and standard deviation as a function of angle and have found no discernible trend or bias.  

Additionally, due to the manner in which we set up the propagating shock, small inhomogeneities arise that are likely the cause of much of the calculated scatter.  This is because we introduce the shock from the boundary conditions which do not explicitly keep the leading edge of the shock as a perfect discontinuity.  Therefore the accuracy of our shock finding algorithm is likely better than that shown in Figure \ref{fig:SA_mach}.

\subsection{Comparison to \citet{Ryu:2003aa}}

\begin{figure*}[tp]
\centering
\includegraphics[width=0.8\textwidth]{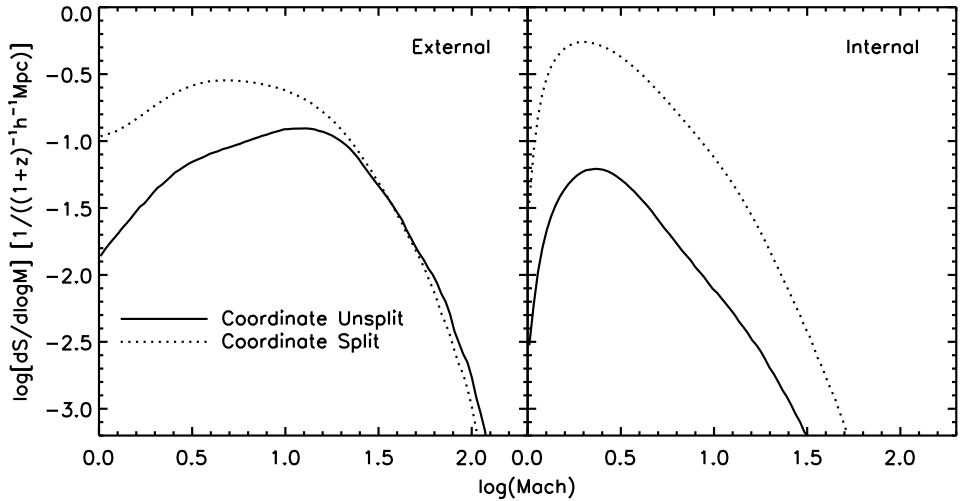}
\caption{Differential shock surface area normalized by the comoving volume of the simulation for external~(left) and internal~(right) shocks for $z=0$ in the ryu1024 simulation.  The two methods of shock finding, coordinate split~(dotted line) and coordinate unsplit~(solid line), are shown.  At low Mach numbers for external shocks and for all internal shocks we see a significant overprediction in the number of shocks when using the coordinate split method described in \citet{Ryu:2003aa}}\label{fig:mach_ryu}
\end{figure*}

In order to test our analysis against previous work done by \citet{Ryu:2003aa}, we generated a $1024^3$ fixed grid simulation with identical cosmological parameters and spatial resolution as their most highly-resolved calculation.  This simulation is dexcribed in detail in Section 2.4.  We expect to see a difference in results from the shock-finding method and from underlying differences in the hydrodynamical solvers.  Enzo uses the Piecewise Parabolic Method, which captures shocks across a single zone, whereas \citet{Ryu:2003aa} use the total variation diminishing (TVD) method, which spreads shocks over approximately two cells.  

In order to study the differences between our shock-finding methods and those of \citet{Ryu:2003aa}, we mimic the top half of Figure 5 from \citet{Ryu:2003aa} in our Figure \ref{fig:mach_ryu}.  However, in addition to using our new shock-finding algorithm that searches along temperature gradients, we include a coordinate-split analysis that is similar to that of \citet{Ryu:2003aa}.  As we claimed in Section 2.2, using a coordinate-split approach overpredicts the number of low Mach number and internal shocks.  For external, low Mach number shocks, the difference is roughly a factor of 3, which agrees with our hypothesis that the coordinate-split approach identifies cells that are associated with a strong shock to the center of a weak shock.  The difference in the internal shocks spans all Mach numbers because the flow to be very complex, making it easy to mistake a normal temperature gradient with that of a shock.  We have also studied the integrated kinetic flux through shock surfaces and find it to be in general agreement with \citet{Ryu:2003aa}.  We will study this in more detail in the future when we include this analysis ``on-the-fly."   

While the shock surface area distributions are good indicators of qualitative differences, we now quantify these results.  This is done by recreating Table 1 from \cite{Ryu:2003aa} in our Table \ref{table:ryucompare}.  For this portion of the analysis, we mimic the Mach number floor requirement that $\Mach >1.5$, which was used to reduce the effects of complex flow in the \citet{Ryu:2003aa} analysis.  First, it is instructive to give a physical motivation for these parameters.  The quantity $1/S$ can be thought of as a mean separation of shocks because it is the simulation volume divided by the total shock surface area.  This gives it units of comoving $Mpc$.  The ratio of external and internal shocks gives the reader an intuition as to where the majority of the shocks are occuring.  Note that as the redshift decreases, the relative amount of internal shocks increases, indicative of the increase in shocks within halos and the measured amount of matter in large halos.  The average quantities are surface area weighted means of the quantity in question.  A subscript of $ext$ or $int$ denotes that only external or internal shocks were used, respectively.  External shocks are those with pre-shock temperatures less than $10^4\ K$ while internal shocks are those with pre-shock temperature greater than $10^4\ K$.

\begin{deluxetable}{ccccccc}[bp]
 \centering    
\tablecolumns{7}
\tablewidth{0pt}
\tablecaption{Mean Shock Quantities}
\tablehead{
 \colhead{z} &  \colhead{1/S} &  \colhead{$S_{ext}/S_{int}$} &   \colhead{$\langle M_{ext}\rangle$} &   \colhead{$\langle M_{int} \rangle$} &  \colhead{$1/S_{ext}$}  & \colhead{$1/S_{int}$}
}
\startdata
          0.0&     6.235&     3.550&      12.65&      3.767&      7.992&      28.37\\
      0.25&     6.519&     4.312&      13.02&      3.961&      8.030&      34.63\\
      0.50&     6.886&     5.196&      12.86&      4.119&      8.211&      42.67\\
      0.75&     7.301&     6.248&      12.61&      4.225&      8.470&      52.92\\
        1.0&     7.767&     7.442&      12.26&      4.310&      8.811&      65.57\\
      1.25&     8.297&     8.743&      11.81&      4.399&      9.246&      80.84\\
      1.50&     8.884&     10.18&      11.34&      4.412&      9.756&      99.38\\
      1.75&     9.546&     11.59&      10.89&      4.411&      10.37&      120.2\\
        2.0&     10.31&     13.04&      10.50&      7.679&      11.10&      144.8 \\ [1ex]
  \enddata
\tablecomments{ Mean shock quantities.  z is the redshift of the simulation.  1/S is the mean comoving length between shock surfaces in units of $h^{-1}Mpc$.  $S_{ext}/S_{int}$ is the ratio of shock surface area for external to internal shocks.    $\langle M_{ext}\rangle$ and $\langle M_{int}\rangle$ are the surface area-weighted mean of the external and internal shock Mach number, respectively.  $1/S_{ext}$ and $1/S_{int}$ are the average comoving distance between external and internal shocks, respectively, in $h^{-1}Mpc$.
}
 \label{table:ryucompare} 
\end{deluxetable}

In comparing our Table \ref{table:ryucompare} to Table 1 in Ryu \etal (2003), we find that we predict a higher average Mach number and higher mean comoving distances between shocks for both internal and external shocks.  The ratio between our average external Mach number and that found in Ryu \etal ranges between 1.54 and 1.6, while that of the internal Mach number~(disregarding $z=2.0$) ranges between 1.3 and 1.5.  We have disregarded $z=2.0$ because there is a large amount of merging between $z=2.0$ and $1.75$, significantly raising the internal temperature of many of the large clusters, increasing the sound speed and decreasing the Mach number.  Therefore, we believe that our particular realization of this volume had later mergers than that of \citet{Ryu:2003aa}. 

The differences in the average Mach numbers as well as the increase in mean comoving distance between shocks is almost entirely due to the use of a coordinate split algorithm vs. a coordinate unsplit algorithm.  The identification of many more low Mach number shocks increases the frequency, thus decreasing the comoving length between shocks.  Therefore for future studies, this difference must be taken into account.

\section{Results for the Santa Fe Light Cone Volume}
\begin{figure*}[tp]
\centering
\includegraphics[width=0.8\textwidth]{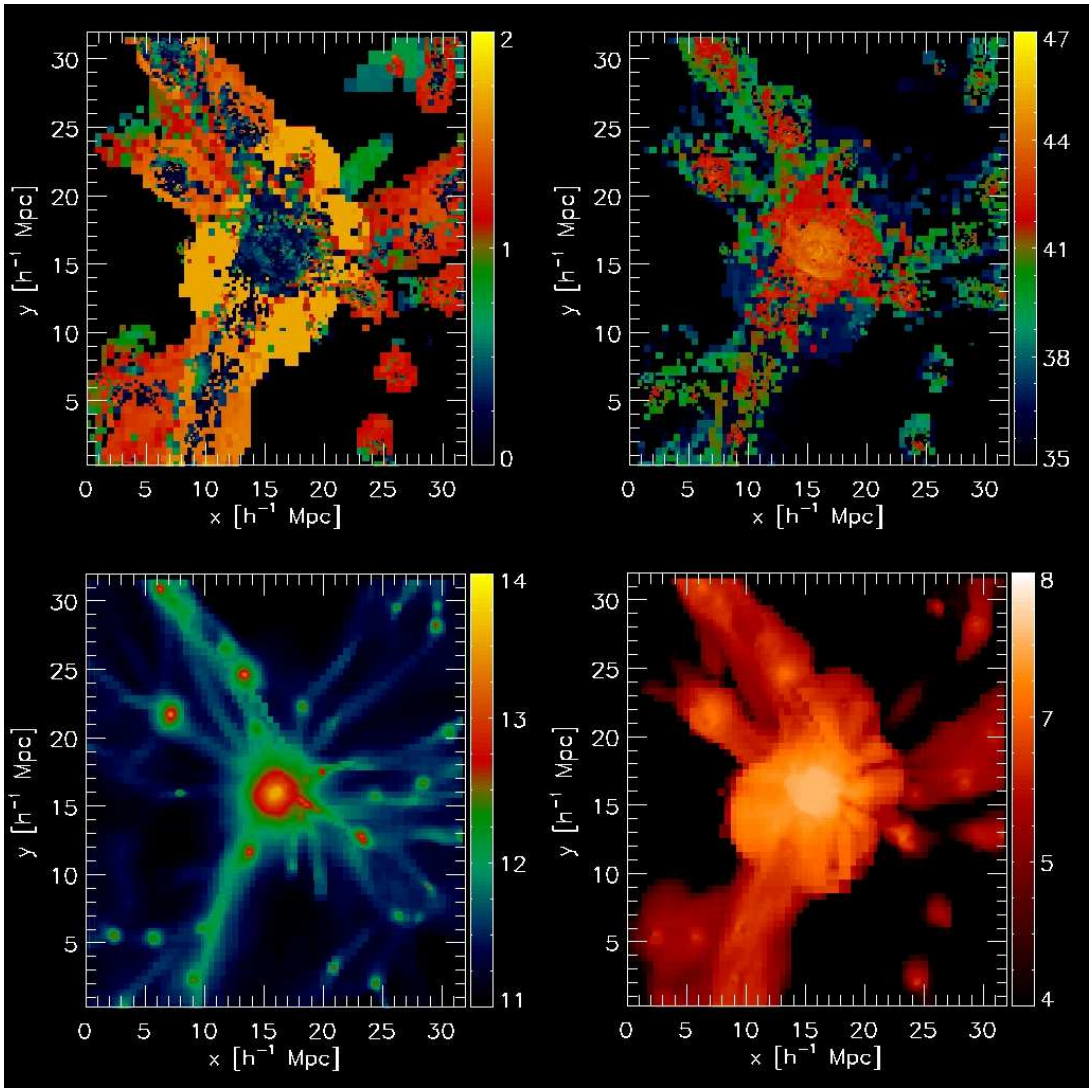}
\caption{Projections of a $2.8\times10^{15}~h^{-1}~M_\odot$ cluster from the ``Santa Fe Lightcone."  Mach number~(top-left) is weighted by the injected cosmic ray flux.  Injected cosmic ray flux~(top-right) is in units of $ergs/(s~h^{-2}~Mpc^{-2})$.  Baryon column density~(bottom-left) is in units of $M_\odot/(h^{-2} Mpc^2)$.  Mass-weighted temperature~(bottom-right) is in units of Kelvin.  The total size of the projected volume is $(32~h^{-1} Mpc)^3$.  All panels show logarithmic quantities.}\label{fig:projections}
\end{figure*}

Now that we have outlined our improved shock finding algorithm, we apply it to a large cosmological simulation encompassing a volume of $\left(512~Mpc/h\right)^3$.  This simulation, called the ``Santa Fe Light Cone,"  was described previously by Hallman \etal (2007).  This represents the first time that a large cosmological volume with superb spatial resolution has been studied for its shock and cosmic ray properties.  Whereas previous studies were only able to study a small number ($\sim10$) of clusters due to a small cosmological box \citep{Pfrommer:2006aa}, we have over 9000 halos with $M_{halo}>5\times10^{13}M_\odot$, and over 200 with $M_{halo}>5\times10^{14}M_\odot$.  This allows us to perform a statistical study of cosmological shocks unlike any that has been done previously.  Both the increase in volume (by a factor of $\sim125$) and an enhanced spatial resolution over previous unigrid/SPH simulations allow unprecedented detail in our calculations.

We begin by outlining the shock distribution and how it can be thought of as a new way to view large scale structure formation in the Universe.  We do this by breaking the distributions down by temperature and density cuts, which further illuminates the underlying dynamics.  From there, we apply the DSA cosmic ray acceleration model and determine what phase of gas will contribute most to the acceleration of cosmic rays.  Finally, we estimate the global fraction of kinetic energy that is processed through shocks that is devoted to the acceleration of cosmic rays in an effort to determine their possible dynamical effects on  gas behavior in galaxy clusters.

\subsection{Shock Frequencies}


As was done in \citet{Ryu:2003aa}, we calculate the surface area of all shocks in a given logarithmic Mach number interval.  However, instead of only classifying shocks as internal or external depending on their preshock temperature, we break the distribution into logarithmic temperature and density cuts that can be postprocessed to examine any subset of the $\rho$ or $T$ phase space for the entire computational volume.  Primarily, we create several physically motivated temperature and density cuts, which are outlined in Table \ref{tab:phase_space}.  Note that the gas in the $T<10^4K$ heading is artificial since we do not include a UV background.  This temperature range traces gas that has not been previously shock heated.  In addition to studying the physical properties of the preshock region, we study the evolution of the distributions as a function of redshift.

Figure \ref{fig:projections} shows a projection of the Mach number for the largest cluster in the simulation $(2.8\times10^{15}M_\odot)$, weighted by cosmic ray acceleration rate.  This allows us to see the structure of cosmological shocks.  By weighting the projection by cosmic ray acceleration rate, we see both the external high-Mach number shocks and the internal shocks, since the internal shocks' weights are higher.  In the other three panels, we show the injected cosmic ray flux, density, and mass-weighted temperature.

\begin{deluxetable}{ccc}
\tablecolumns{3}
\tablewidth{0pt}
\tablecaption{Temperature-Density Phase Space}
\tablehead{
\colhead{Location} & \colhead{Temperature Range} & \colhead{Overdensity Range}
}
\startdata
 	Voids&				$T\ <\ 10^4K$&   			$\delta <1$\\
	Filaments&			$10^4K\ <\ T\ <\ 10^6K$&   	$1\ <\ \delta <100$\\
	Clusters&				$10^6K\ <\ T\ <\ 10^8K$&   	$100\ <\ \delta <10^3$\\
	Cluster Cores&	$T\ >\ 10^8K$&   			$\delta\ >\ 10^3$      \\[1ex]
\enddata
\tablecomments{Approximate ranges for pre-shock temperature $T$ or pre-shock overdensity $\delta=\rho_b/\langle\rho_b\rangle$ for general large scale structures.
}
  \label{tab:phase_space}  					
\end{deluxetable}

\subsubsection{Density \& Temperature Ranges}

\begin{figure*}[tp]
\centering
\includegraphics[width=0.9\textwidth]{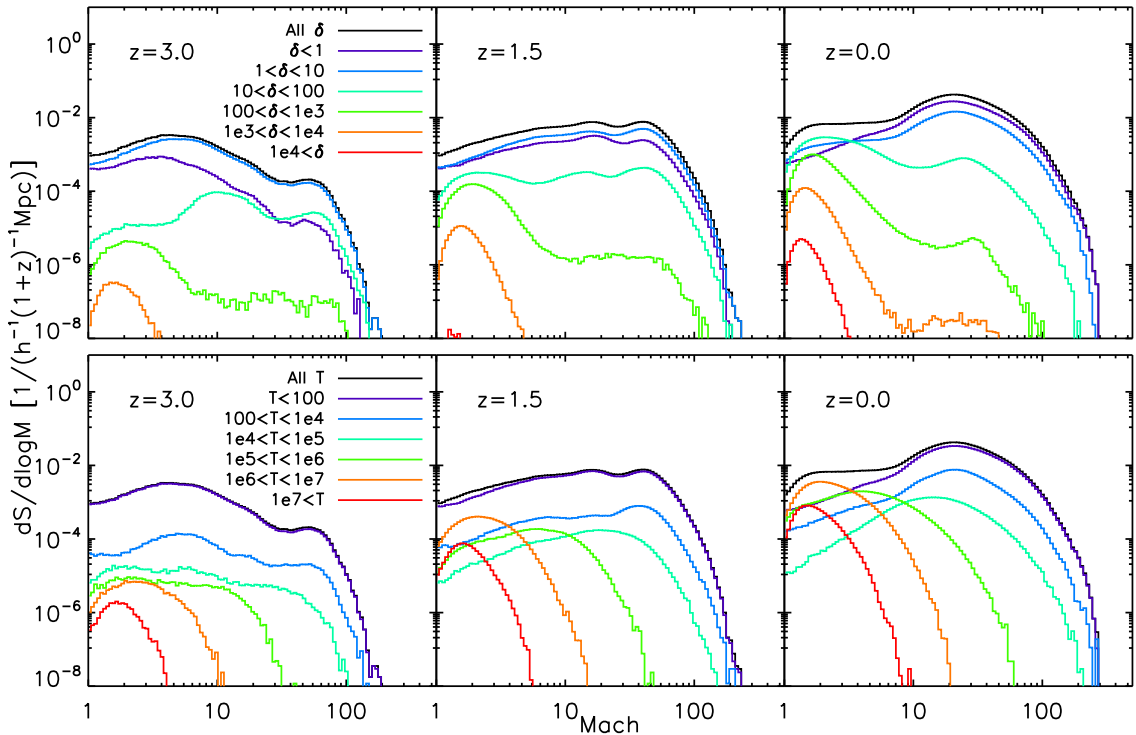}
\caption{Differential shock surface area as a function of logarithmic mach number bins for varying pre-shock gas phases.  Pre-shock gas overdensity~(top) is divided into several ranges that differentiate the overall distribution.  Pre-shock gas temperature~(bottom) differentiates the different types of shocks (i.e. accretion, merger).  Both distributions are shown for three redshifts: $z=3$~(left), $z=1.5$~(middle) and $z=0$~(right).}\label{fig:mach_sw_tempden}
\end{figure*}


We now expand the classification of external and internal shocks \citet{Ryu:2003aa} by examining the shock Mach number distributions in varying temperature and density ranges.  This will provide a more complete description of where these shocks arise in structure formation than in previously published analyses.  In Figure \ref{fig:mach_sw_tempden}, the shock frequency is plotted for a range of density and temperature cuts.  At $z=3$, we see that shock surface area distribution is dominated by shocks with low temperature/low overdensity pre-shock quantities.  These represent the accretion shocks onto filaments and proto-clusters.  As the simulation evolves, the distribution becomes bimodal with components from both low pre-shock temperature, high-Mach number accretion shocks and high-temperature, low-Mach number merger shocks.  

The temperature cuts each have a characteristic Mach number cutoff that increases with decreasing temperature.  This cutoff is due to the maximum temperature jump that is possible with a given pre-shock temperature.  Therefore, since the maximum temperature in the simulation is $\sim10^8K$(determined by the mass of the largest cluster), a temperature jump from $10^6K$ will result in a $\Mach \approx 18$, very close to the cutoff seen at $z=0$ for $10^6K<T<10^7K$.  Similarly, the cutoffs for lower pre-shock temperatures indicate the largest temperature jumps for each population.  At higher redshifts, these temperature cutoffs decrease due to the lower maximum temperatures present in the simulation.  Therefore, the movement of these cutoffs tell us about the temperature evolution of the simulation.

Additionally, the Mach number associated with the peaks in the shock frequency distribution can be used to determine the mathematical mode of the post-shock temperature distribution.  For $z=0$, these peaks correspond to post-shock regions with $T_2\sim$few$\times10^6K$ for pre-shock temperatures $T_1 < 10^6K$.  Therefore the majority of these shocks are heating the pre-shock gas to WHIM temperatures in filaments.  For $T_1>10^6K$, the peak Mach numbers correspond to post-shock temperatures of $T_2\approx 2\times10^7 - 1.5\times10^8K$.  These are complex flow and subhalo merger shocks in the interior regions of clusters.  

If we instead examine the varying density cuts, similar results are observed.  At high redshifts, we see that the dominating accretion shocks (high Mach number shocks) have pre-shock overdensities of $\delta\sim1-10$.  This is because of the relative paucity of large-scale halos and filaments and, thus, relatively shallow gravitational potential wells.  The infalling gas will get much closer to the accretor and therefore denser before shocking.  As we move to lower redshift, the $\delta<1$ shocks begin to dominate because we are shocking further out into the voids.  

For the interior cluster shocks, there are three regimes that present themselves in the analysis.  If we examine $z=3$ with $10< \delta<100$, there are plateaus near $\Mach\approx 2-4$ and $\Mach\approx10-70$.  It is difficult to determine what the post-shock density will be because of the insensitivity of the density contrast at high Mach numbers ($\rho_2/\rho_1 \rightarrow 4$ for $\Mach >>3$).  However, it is likely that the two high Mach number shock plateaus correspond to filaments for $\Mach\approx10$ and clusters at the virial radius for $\Mach\approx70$.  The low Mach number shocks are most likely interior flow shocks.

At late times, all of the intermediate pre-shock density regions have bimodal distributions.  The high Mach peak corresponds to density contrasts of 4, while the low $\Mach$ corresponds to jumps of $\sim 2$.  Therefore, we are likely looking at merger and complex flow shocks, respectively.


\subsubsection{Redshift Evolution of Shock Properties} 

\begin{figure}[t]
\centering
\includegraphics[width=0.4\textwidth]{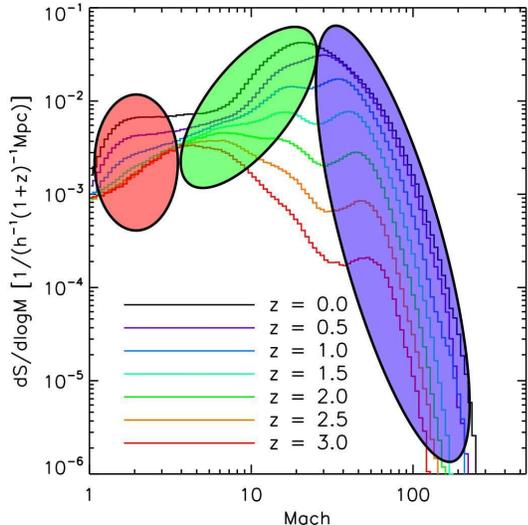}
\caption{Comoving shock surface area normalized by the simulation volume  as a function of Mach number with varying redshifts.  Three regions are suggested corresponding to internal cluster merger shocks~(red), accretion shocks onto filaments~(green), and accretion shocks onto clusters~(blue).}\label{fig:mach_sw_red}
\end{figure}

There are three primary populations of shocks that we see evolve through time, as seen in Figure \ref{fig:mach_sw_red}.  There are accretion shocks onto clusters, accretion shocks onto filaments, and merger and complex flow shocks within clusters and filaments.  These are outlined by the blue, green, and red shadings in Figure \ref{fig:mach_sw_red}, and their qualitative behavior can give useful insights as to the evolution of large scale structure.  Let us analyze each of these populations separately.  To determine the origin of these populations, we have examined slices and projections of the data and compared the Mach number of the cell to it's location with respect to large scale structure.

First, at early times we see a small peak at very high Mach numbers that denotes shocks onto collapsing halos.  This corresponds to gas that has previously been untouched by shocks falling directly onto the proto-cluster gas, with temperature jumps from hundreds of Kelvin to $10^6K$ (Note the Mach numbers are still calculated with a temperature floor of $10^4K$).  We see that as the universe evolves, the strongest shocks in the simulation become stronger.  This is due to the mass of the clusters increasing with time, providing a larger gravitational force pulling the material onto the halo.  We also see that this peak increases in shock frequency while slowly moving to slightly lower Mach numbers.  Because the mass function cuts off exponentially at high mass, the number of small halos heavily outweighs the large halos.  These smaller halos have lower free-fall speeds at the radius of the accretion shock, leading to a smaller Mach number.  Therefore the large number of weaker shocks dominate the net surface area distribution.

Second, the shocks onto filaments begin at Mach numbers of $\Mach \sim 6$ and move towards $\Mach \sim 20$ at late times.  The surface area of these shocks are much larger at early times because the surface area of a cylinder per unit volume is larger than that of a sphere as well as an increased number of filaments with respect to halos (there are several filaments that feed into a single halo).  The strength of these shocks grow with the increase in size of the filaments.  

Finally, the low Mach number shocks ($\Mach < 3$) due to halo mergers and complex flow are nearly non-existent at high redshifts.  However, as large halos collapse and start to merge, the shock surface area also increases.  Therefore this evolution traces the strength and frequency of merger shocks.

\subsection{Cosmic Ray Energy Injection}

The thermal history of the large scale structure in the Universe is primarily determined by the conversion of gravitational potential energy into kinetic energy, which is subsequently converted to heating gas and the acceleration of cosmic rays.  Here we present results of our application of the cosmic ray acceleration model described in Section 2.3 to the ``Santa Fe Light Cone."

\subsubsection{Function of Redshift}

\begin{figure}[t]
\centering
\includegraphics[width=0.4\textwidth]{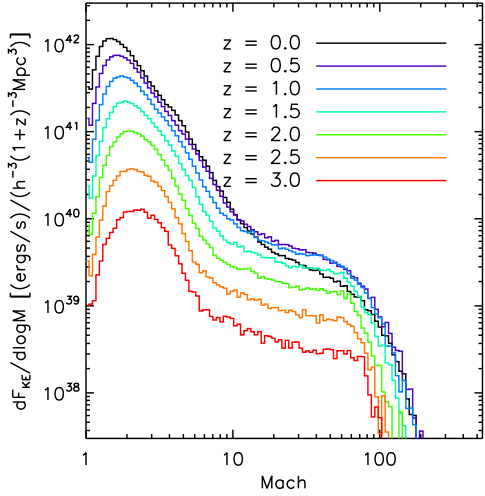}
\caption{Redshift evolution of the amount of kinetic energy processed by shocks as a function of Mach number.  Redshift decreases from $z=3$~(red) to $z=0$~(black).  The decrease in flux at late times for $\Mach>10$ signals the epoch at which dark energy becomes dominant. }\label{fig:kin_flux}
\end{figure}

\begin{figure}[t]
\centering
\includegraphics[width=0.4\textwidth]{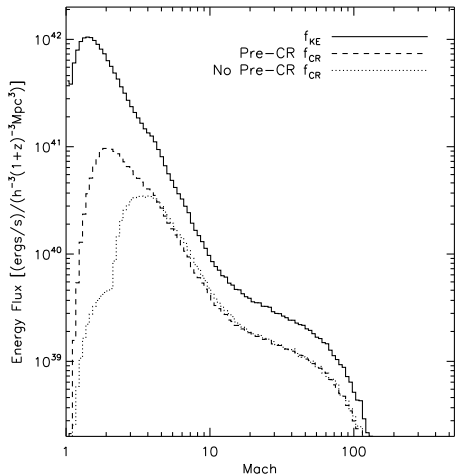}
\caption{Out of the incoming total kinetic energy of the shocks (solid line), the relative amount of energy devoted to the acceleration of cosmic rays for both models with~(dashed line) and without~(dotted line) a pre-existing CR population, as predicted by the \citet{Kang:2007aa} diffusive shock acceleration model.}\label{fig:energy_breakdown}
\end{figure}

The first result is that, as in previous studies \citep[e.g][]{Ryu:2003aa, Pfrommer:2006aa, Kang:2007ab}, the most important Mach number shocks in terms of cosmic ray acceleration are at $\Mach \approx2-4$.  This may seem surprising given that the surface area of shocks is dominated by high Mach shocks.  However, the amount of energy dissipation is the product of the mass flux through the shocks and the Mach number.  The large accretion shocks at early times ($z<3$) have already consumed a large fraction of the gas in voids.  This leaves very little mass at low densities to be processed by the most massive halos.  This is in contrast to the low Mach complex flow shocks within the clusters.  These process very large amounts of mass and kinetic energy, and therefore experience very high thermalization and acceleration of cosmic rays even with lower efficiency.  Cosmic rays from these low Mach number shocks will, however, have a steep energy spectrum and dissipate their energy relatively quickly compared to strong accretion shocks \citep[e.g.][]{Miniati:2001aa}.

Figure \ref{fig:kin_flux} shows a distribution function of the kinetic energy processed through shocks per comoving $(Mpc/h)^3$ as a function of redshift where the height of the distribution function gives the differential amount of kinetic energy processed by shock for a given Mach number bin.  As the simulation evolves to $z=0.5$, there is a monotonic increase in the average kinetic energy density processed.  Both the low-Mach complex flow and high-Mach accretion shocks increase by factors of $10-100$.  This monotonic increase stops at $z\sim0.5$ because of the dominance of dark energy in a  $\Lambda CDM$ universe at this epoch, resulting in a decreased merger of, and accretion onto, the highest-mass halos.  Therefore, the number of accretion shocks characterized by high Mach numbers will decrease.  Compounding this effect is the slow evacuation of the voids and the lack of additional mass to accrete.  

By applying the diffusive shock acceleration model, we can estimate how much of this energy is put into gas thermalization versus the acceleration of cosmic rays within the confines of the model.  This acts
as a first estimate of the energy injection into cosmic rays, and should not be taken as the final on the subject.  Cosmic ray injection is a highly non-linear process that is not fully understood.  Further work on
this model is needed.

Figure \ref{fig:energy_breakdown} shows the relative amounts of energy dissipated for the two different models involving either no pre-existing cosmic rays or an initial amount of cosmic rays such that $P_{CR}/P_{g} \approx 0.3$.  As one can see, the relative amount of cosmic ray acceleration vs. thermalization heavily depends on the assumed inputs of the underlying DSA model.  Until we are able to track the cosmic ray pressure within our simulations, we are resigned to give these rough limits of cosmic ray acceleration.

\begin{figure*}[t]
\centering
\includegraphics[width=0.9\textwidth]{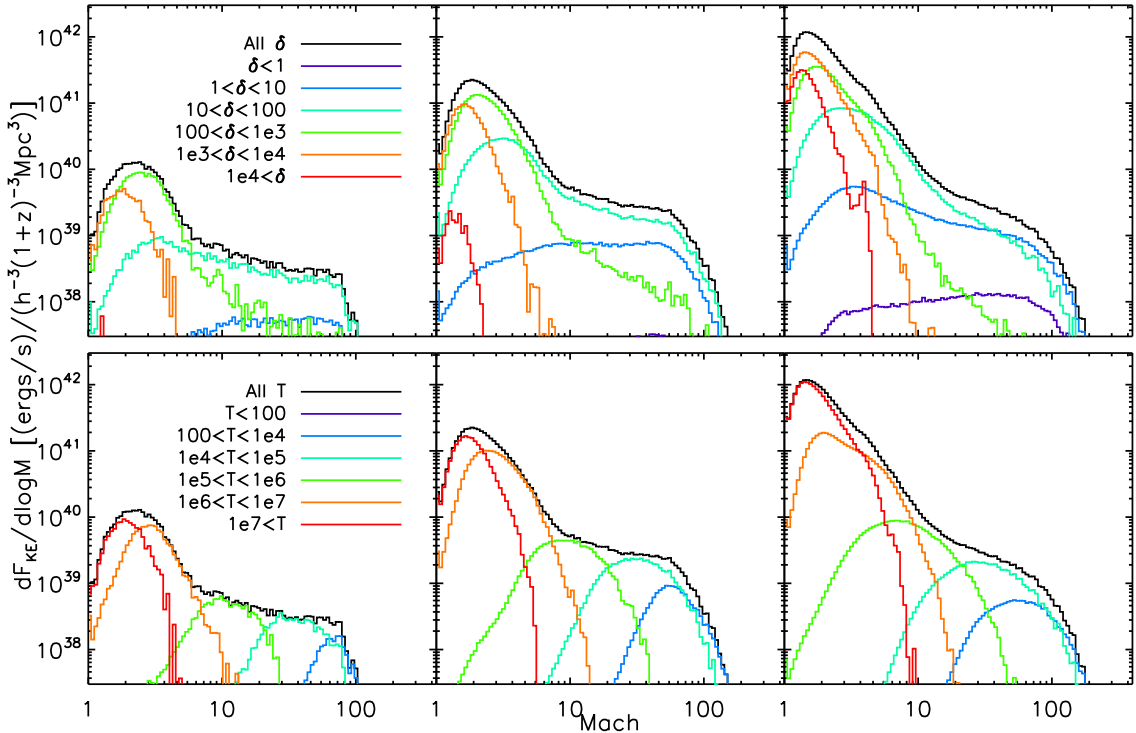}
\caption{Differential kinetic energy flux processed by shocks as a function of Mach number and pre-shock gas phase.  Pre-shock gas overdensity~(top row) is divided into several ranges that differentiate the overall distribution.  Each pre-shock temperature range~(bottom row) roughly corresponds to a particular mach number.  Both distributions are shown for three redshifts of $z=3$~(left column), $z=1.5$~(middle column) and $z=0$~(right column).}\label{fig:mach_thermw_temp}
\end{figure*}

\subsubsection{Variation of Cosmic Ray Injection Efficiencies With Gas Properties}

Separation of distribution functions showing thermalization as a function of both temperature and density provides valuable insight into the physical processes occurring in the simulation.  In Figure \ref{fig:mach_thermw_temp}, we see that there are two primary modes of kinetic energy flux at $z=3$.  For $\Mach<2$, the thermalization is dominated by shocks at $100 \lesssim \delta \lesssim 10^4$ and $T \gtrsim 10^6K$.  These shocks are likely within the largest filaments and the first clusters.  At higher Mach numbers, $\Mach > 6-7$, the thermalization is dominated by gas at $T<10^6 K$ and $\delta \sim 10-100$.  This points towards accretion shocks onto filaments and the heating of the WHIM.  If we use the peaks in each temperature cut up to $T\sim10^6K$ to estimate the Mach number, we can calculate the post-shock temperature for these shocks to be $1-3\times10^6K$.  This reinforces the thought that these shocks are heating the WHIM.  Shocks in this range of Mach numbers are also seen in Figure \ref{fig:projections} as surrounding the filaments.  

At later times, the entire distribution shifts to higher thermalization rates due to the collapse of structures.  Low Mach numbers are again dominated by complex flows within clusters.   By examining the shocks with pre-shock temperature of less than $10^5K$ as well as the redshift evolution from Figure \ref{fig:kin_flux}, we are able to verify that the high Mach number accretion shocks are becoming less important due to the separation of the voids from the clusters after $z\approx0.5$.  In the overdensity cut that corresponds to $1<\delta<10$, we see a shift from a peak at high Mach numbers to small Mach numbers as the relative importance of accretion and mergers switch.  

If we compare our results to those of \citet{Pfrommer:2006aa}, we see a good agreement at low Mach numbers.  \citet{Pfrommer:2006aa} found shocks as strong as $\Mach \sim 10^3$.  However, we never see shocks above $\Mach\approx 200$.  This is likely due to the lack of a temperature floor in their simulation, which thus allows a higher numerical value for the Mach number.  These shocks are likely not realized in the real universe due to the presence of a ubiquitous ionizing radiation background that will keep gas above $10^4K$.  

\begin{figure}[t]
\centering
\includegraphics[width=3in]{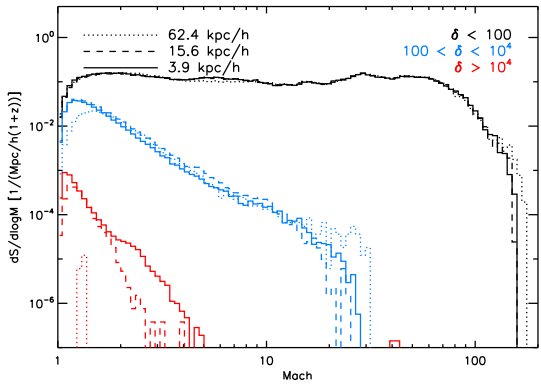}
\caption{The effects of spatial resolution in different density regimes.  Here we keep the mass resolution at the highest level ($M_{dm} = 9.7\times10^8 M_\odot$) .  The distribution function of shock Mach numbers in a $(32~Mpc/h)^3$ volume around a cluster weighted by surface area for three overdensity cuts is plotted against Mach number.  Three cuts in overdensity are shown for $\delta<100$~(black lines), $100<\delta<10^4$~(blue lines),$\delta>10^4$~(red lines).  Varying spatial resolution are shown with dotted ($62.4~kpc/h$), dashed ($15.6~kpc/h$), and solid ($3.9~kpc/h$) linestyles.}\label{fig:density_level}
\end{figure}

Finally, we can compare our results to recent work by Kang \etal (2007), who used a unigrid calculation similar to that of \citet{Ryu:2003aa}, but included radiative processes, star formation, and a relaxed minimum temperature floor.  Again, this relaxation of the temperature floor to (in their case) the CMB temperature resulted in very high Mach numbers -- up to $\Mach>10^4$.  This corresponds to a temperature jump by a factor of $\sim 3\times 10^7$, a jump from $3K$ to $10^8K$ (the minimum and maximum temperatures in the simulation).  At low Mach numbers, our results are very similar to those of \citet{Kang:2007aa}.  

\section{Effect of Mass and Spatial Resolution on Cosmic Ray Acceleration Efficiency}
In order to quantify the robustness of our simulations with respect to mass and spatial resolution, we perform a series of simulations where the mass and spatial resolution of a single galaxy cluster are varied over a wide range of parameter space.  Two parameters are varied in this study.  The first is the maximum level of refinement, which affects the spatial resolution and, ultimately, the accuracy of the hydrodynamic solver.  The second parameter is the dark matter particle mass resolution, which affects the accuracy with which the gravitational potential is calculated.

\subsection{Spatial Resolution}

\begin{figure}[tp]
\centering
\includegraphics[width=3in]{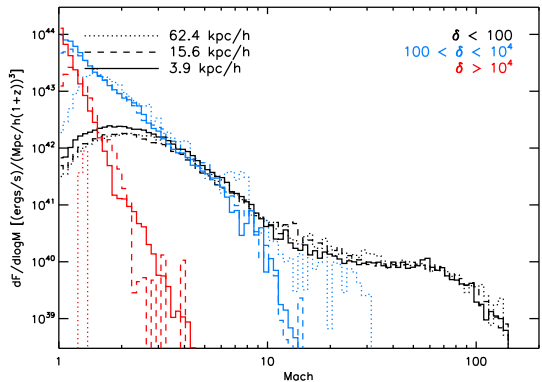}
\caption{Spatial resolution effects on kinetic energy flux for several density regimes.  The kinetic energy flux is plotted against Mach number for varying spatial resolution denoted by linestyle.  The shocks are grouped as external (black lines), clusters/filaments (blue lines), and rich clusters (red lines).}\label{fig:therm_diss_level}
\end{figure}

Our maximum spatial resolution ranges from $62.4~kpc/h$ to $3.9~kpc/h$ (see Table \ref{tab:simparams}).  Since this only limits the maximum resolution, one expects to see a strong dependence on this parameter only at high densities.  Figure \ref{fig:density_level} shows the dependence of shock surface area on level refinement for three overdensities.  For shocks with pre-shock overdensities less than $\sim100$, the main difference in the multiple resolutions is at low Mach numbers (below $M\sim2$) and only appears in the lowest resolution simulation.  

At overdensities above $10^3$, we not only see that the low Mach number complex flow shocks are lost at low resolution, but also a drop in the number of high Mach number shocks.  At this density and spatial refinement, it is thought that the absence of sufficient spatial resolution leads to the artificial smoothing of the gas, creating an inability to capture shocks.  The main result of this spatial resolution study is that a resolution between $3.9~kpc/h$ and $15.6~kpc/h$ should be sufficient in all but the most dense regions of the simulations.  Therefore, our ``Santa Fe Light Cone" simulation presented in Section 3 had an adequate peak resolution of $7.8~kpc/h$.

To examine the effect of spatial resolution on cosmic ray acceleration, we study the kinetic energy flux through shocks as a function of spatial resolution.  Figure \ref{fig:therm_diss_level} shows a distribution function measuring the thermal dissipation rate as a function of Mach number with varying spatial resolution.  This study is performed with the maximum mass resolution, $M_{dm}=9.7\times10^8M_\odot$.  At low overdensities ($\delta<100$), the effect of spatial resolution is very small.  At moderate to high overdensities ($100<\delta<10^4$), there are differences on the order of a factor of $~2$ that are likely due to the smoothing of high density gas  as the resolution is decreased.  The primary difference in the dissipation rates occur for low Mach numbers when we do not have sufficient spatial resolution to resolve all of the complex flow shocks.  There are also large differences at $\Mach \geq 10$ for the lowest resolution simulation.  However, the difference between $3.9\ kpc/h$ and $15.6\ kpc/h$ is negligible.

At very high overdensities ($\delta > 10^4$), there is a very large difference between the varying spatial resolutions.  One reason is that if a cell has an overdensity of $10^4$, the grid would normally be on the 5th level of refinement.  With a maximum refinement level of 4 for the poorest resolution simulation, any gas at this overdensity would be very poorly resolved.  The difference between the  $15.6~kpc/h$ and $3.9~kpc/h$ resolution simulations is likely small number statistics for the former simulation.  The $3.9~kpc/h$ resolution simulation will resolve these high densities with roughly 64 times more cells compared to the $15.6~kpc/h$ simulation.  

\begin{figure}[tp]
\centering
\includegraphics[width=3in]{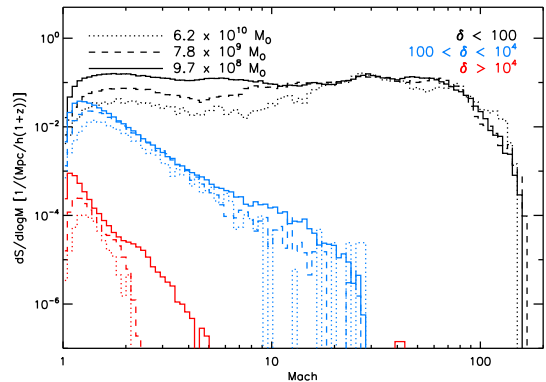}
\caption{The effects of mass resolution in different density regimes.  Here we keep the spatial resolution at the highest level of $3.9 kpc/h$.  Mass resolutions are shown by linestyles of dotted ($6.2\times10^{10}M_\odot$), dashed ($7.8\times10^{9}M_\odot$), and solid ($9.7\times10^{8}M_\odot$).  The shocks are grouped as external (black lines), clusters/filaments (blue lines), and rich clusters (red lines).}
\label{fig:density_mass}
\end{figure}   



\subsection{Mass Resolution}

The mass resolution of each simulation is set by the resolution of the root grid (or highest-level static nested grid).  The size of each root cell determines the amount of mass given to each dark matter particle.  Therefore, if the root grid doubles in resolution, the mass resolution increases by a factor of $2^3$.  In principle, there should be two effects of increased mass resolution.  First, one would expect that since we are extending our mass function to a lower limit, the number of subhalos and our resolution of  complex fluid flow should increase.  This should manifest itself in an increase of shocks in the low Mach number regime.  Second, the increased mass resolution also corresponds to an increase in the static grid spatial resolution.  This may affect the calculated surface area of shocks that reside in voids.  Since the temperature jumps in the voids are likely to be much higher than those within clusters, we would expect this effect to show up in the high Mach number regime.

In order to test these hypotheses, we varied the mass resolution from $6.2 \times 10^{10} M_{\odot}/h$ to $9.7 \times 10^8 M_{\odot}/h$.  The results of this study are shown in Figure \ref{fig:density_mass}.  At $\delta<100$, we see that as the mass resolution increases, the number of low Mach number shocks increase, while the high Mach number shocks decrease.  At high densities, the situation is more complicated.  For $\Mach< 2$, the surface area likely increases because of the increase in the number of subhalos and complex flow.  For $M>7-8$, the differences seem to be largely due to statistical uncertainties.  For $\delta>10^4$, the disparity at $\Mach<2$ is again likely due to the number of subhalos and their effects on turbulence.  At $2<M<5$, there is a large difference between the highest mass resolution simulation and the other two.  Because we believe these shocks are merger shocks, it may be because there are just too few dense subhalos that merge with large halos to create this surface area.

\begin{figure}[tp]
\centering
\includegraphics[width=3in]{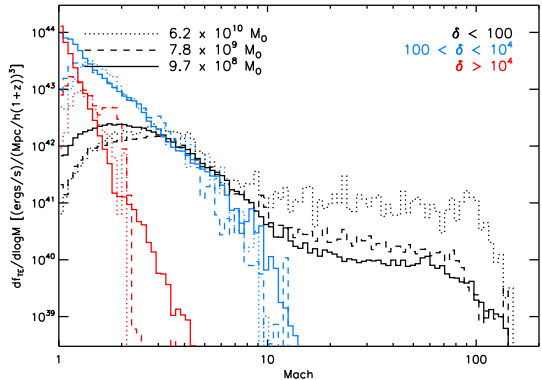}
\caption{Spatial resolution effects on kinetic energy flux for varying dark matter particle masses. Mass resolutions are shown by linestyles of dotted ($6.2\times10^{10}M_\odot$), dashed ($7.8\times10^{9}M_\odot$), and solid ($9.7\times10^{8}M_\odot$).  The shocks are grouped as external (black lines), clusters/filaments (blue lines), and rich clusters (red lines).}\label{fig:therm_diss_mass}
\end{figure}

As with the spatial resolution, we now study the effects of mass resolution on the thermal dissipation rates at shock fronts.  Figure \ref{fig:therm_diss_mass} shows the effect of varied mass resolutions with a fixed spatial resolution of $3.9~kpc/h$.  Again, we break the analysis down into overdensity regimes.  Low overdensity, high Mach number shocks exhibit a strong dependence on the mass resolution.  This is because of the ability to better resolve shocks in the voids and low density filaments.  The disparity in high overdensity ($100<\delta<10^4$), low Mach number shocks is less apparent, but also suggests that the mass resolution of the simulation has an effect on the thermal dissipation of gas through shocks.  At $\delta>10^4$, we again see the effect of a decreased number of subhalos available to merge.

Contrary to the effects of spatial resolution, the biggest differences due to mass resolution appear in the high Mach number regime.  This is again due to the overestimate of Mach number at low root grid resolution.  One evident result is that while the spatial resolution seems to be fairly well converged, it is not clear that the mass resolution has converged.  Therefore, we can only claim a fairly weak precision in the thermal dissipation and cosmic ray acceleration rates for the current simulations.

There are several key results to this resolution study.  We appear to have converged in terms of maximum spatial resolution in all but the densest cluster gas.  However, our convergence upon the various quantities with respect to dark matter mass resolution is not clear.  The differences in the cosmic ray acceleration rate are not larger than the underlying uncertainty in the results of diffusive shock acceleration simulations, suggesting that both mass resolution and our understanding of the physical mechanisms of cosmic ray acceleration must be improved in the future.


\section{Discussion}

There are several topics that warrant discussion with respect to the results that we have presented thus far.  These include the variation of results with respect to $\sigma_8$, the inclusion of non-adiabatic physics, the limitation of the diffusive shock acceleration model, and the implications of the mass resolution in the ``Santa Fe Light Cone" Simulation.

If our goal is to do large statistical studies of galaxy clusters, changing the value of $\sigma_8$ will have significant effects.  First, a higher $\sigma_8$ will greatly increase the number of massive clusters in a given volume.  By comparing the $ryu1024$ simulation with the ``Santa Fe Light Cone," with values of $\sigma_8$ of 0.8 and 0.9, respectively, we see that this increases the frequency and strength of the high Mach number shocks.  Additionally, this should increase the amount of kinetic energy that is processed by shocks since mergers will be more frequent.  

In all of our simulations thus far we have only used adiabatic physics.  Previous studies, such as those done by \citet{Kang:2007ab}, have found that when including radiative cooling and star formation that the shock properties are still governed primarily by gravitational physics and that additional physics have little effect on overall distributions at scales larger than $\sim 100 h^{-1} kpc$.  However, \citet{Pfrommer:2007aa} found that at smaller scales, on the inside of clusters, the cosmic ray contribution to the overall pressure is greatly increased with the inclusion of radiative cooling.  Additionally, we currently adopt a temperature floor of $10^4K$ because of the lack of an ionizing background.  This should instead be done in a self-consistent manner.

While we are using results of recent diffusive shock acceleration simulations by \citet{Kang:2007aa}, there are assumptions and limitations that may have an effect on our results.  We assume that the magnetic field is parallel to the shock normal, which yields the largest efficiency for accelerating cosmic rays.  Any deviation from this will likely cause decreases in the overall efficiency of the shocks as particle accelerators.  Additionally, for low Mach numbers, knowledge of the pre-shock composition is very important and can lead to orders of magnitude differences in the acceleration efficiency.  Therefore, being able to track the cosmic ray pressure in ``on-the-fly" calculations will allow us to provide a more self-consistent estimate.  Finally, we are assuming that the only method for cosmic ray production is through first-order Fermi acceleration, and therefore we ignore other potetial sources of cosmic rays, such as second-order acceleration by turbulence, galaxies, and AGN.  

The results of the resolution study provided in Section 4 have suggested that we have not yet seen a convergence with respect to the dark matter particle mass in the ``Santa Fe Light Cone" simulation.  This likely results in an under-prediction in the number of merging subhalos and the kinetic energy flux associated with them.


\section{Conclusions \& Future Directions}

Our study of cosmological shocks has resulted in several advances in both scientific understanding and numerical algorithms.  We now summarize the key findings: 
\begin{itemize}
\item We have developed a novel numerical scheme that is capable of detecting and accurately characterizing the Mach number of shocks in an adaptive mesh refinement simulation.  This method has relaxed the previous restriction of using a coordinated axis-based approach and now allows us to accurately characterize shocks that have any orientation with respect to the coordinate grid.

\item Using our new shock-finding technique on a unigrid cosmological simulation that is identical to the highest-resolution calculation in \citet{Ryu:2003aa}, we have shown that previous methods resulted in an overestimate of the number low Mach number shocks by a factor of $\sim3$ due to confusion of the direction of shock propagation, and that this underestimate is consistent with using shock-finding algorithms that only sweep along coordinate axes.  

\item We have analyzed the largest AMR cosmological simulation to date that includes adiabatic gas physics, the ``Santa Fe Light Cone."  This simulation has an effective spatial dynamic range of $65,536$ and resolves both large scale structure and small-scale features within galaxy clusters.  Whereas previous studies were able to study on the order of 10 high mass clusters, we have thousands within a single simulation volume.  Our study of this simulation has led to a new technique for conceptualizing structure formation because we are able to analyze the evolution of three different populations of shocks: cluster accretion, filament accretion, and internal merger and complex flow shocks.  

\item  By applying the results of 1-D Diffusive Shock Acceleration models, we calculate the amount of kinetic energy at shock fronts that is used to accelerate cosmic rays, and find it to be in agreement with previous studies.  These cosmic rays will make up a significant fraction of the total pressure in the intracluster medium and therefore their dynamical effects need to be studied.

\item We have performed a resolution study that varies both the dark matter particle mass and peak spatial resolution.  From the results of this study, we believe that the spatial refinement in the ``Santa Fe Light Cone" simulation is adequate.  The state of mass resolution convergence is less clear, suggesting that for future studies we should focus on higher mass resolution.

\end{itemize}

While our numerical technique of characterizing shocks has been proven to be robust, our results are still somewhat limited by the physics.  We have not yet included potentially important effects such as radiative cooling, star formation and feedback, AGN heating, or a photoionizing UV background.  These physics will be included in future work.  Second, our results are based upon a post-processing of the simulation output.  Ideally, the shocks would be identified in an ``on-the-fly" manner during simulation runtime.  Additionally, the cosmic ray acceleration would be traced in a self-consistent manner that allowed for a back-reaction on the gas.  Attempts at tracing the cosmic ray pressure have been made by \citet{Pfrommer:2006aa,Pfrommer:2007aa} using an SPH code, and we will be working towards the same goal in the near future within $Enzo$.  Finally, the acceleration of cosmic rays is still dependent on the underlying magnetic field strength and orientation.  Cosmological MHD has been implemented within $Enzo$, and in the near future we will include magnetic fields and their coupling to cosmic rays within a cosmological AMR volume.  

\acknowledgments{We would like to thank Hyesung Kang and Tom Jones for making their DSA model fits available.  We would also like to thank Dongsu Ryu, Christoph Pfrommer, Greg Bryan, Tom Jones, and an anonymous referee for useful comments.  B.W.O. would like to thank Greg Bryan, Stirling Colgate, Hui Li, and Franco Vazza for useful discussions.  B.W.O. carried out this work under the auspices of the U.S. Department of Energy at Los Alamos National Laboratory under Contract no. DE-AC52-06NA25396.  This work has been released under LA-UR-07-7916.  Simulations and data analysis were performed at SDSC and NCSA with computing time provided by NRAC allocation MCA98N020 and at LANL under the Institutional Computing program.  S.W.S., E.J.H. and J.O.B. have been supported in part by a grant from the US National Science Foundation (AST 04-07368). E.J.H. also acknowledges support from NSF AAPF AST 07-02923.} 

\vspace{5mm}

\bibliography{ms}

\end{document}